\documentclass
[doublecol]
{epl2}
\usepackage{graphicx}  % needed for figures
\usepackage{dcolumn}   % needed for some tables
\usepackage{bm}        % for math
\usepackage{amssymb}   % for math
\usepackage{amsmath}
\usepackage{color}

\usepackage[T1]{fontenc}
\usepackage[latin9]{inputenc}
\usepackage{amsthm}
\usepackage{amsmath}
\usepackage{graphicx}
\usepackage{xcolor}

\usepackage{subcaption}

\makeatletter
\@ifundefined{textcolor}{}
{%
 \definecolor{BLACK}{gray}{0}
 \definecolor{WHITE}{gray}{1}
 \definecolor{RED}{rgb}{1,0,0}
 \definecolor{GREEN}{rgb}{0,1,0}
 \definecolor{BLUE}{rgb}{0,0,1}
 \definecolor{CYAN}{cmyk}{1,0,0,0}
 \definecolor{MAGENTA}{cmyk}{0,1,0,0}
 \definecolor{YELLOW}{cmyk}{0,0,1,0}
 }

\makeatletter

\newcommand{\Rmnum}[1]{\expandafter\@slowromancap\romannumeral #1@}
\makeatother

\newcommand{\be}{\begin{equation}}
\newcommand{\ee}{\end{equation}}

\def\lsim{\mathrel{\rlap{\lower4pt\hbox{$\sim$}}
    \raise1pt\hbox{$<$}}}                % less than or approx. symbol
\def\gsim{\mathrel{\rlap{\lower4pt\hbox{$\sim$}}
    \raise1pt\hbox{$>$}}}

\renewcommand\[{\begin{equation}}
\renewcommand\]{\end{equation}}
\usepackage[english]{babel}
\addto\captionsenglish{\renewcommand{\figurename}{FIG.}}
\makeatother
\newcommand*\samethanks[1][\value{footnote}]{\footnotemark[#1]}
\usepackage{babel}

\begin{document}

\title{Dead or alive: Distinguishing active from passive particles using supervised learning}
\author{Giulia Janzen\inst{1,2} \thanks{These authors contributed equally to this work.}\and Xander L.J.A.~Smeets\inst{1} \samethanks \and Vincent E.~Debets\inst{1,2} \and Chengjie~Luo\inst{1,2}\and Cornelis~Storm\inst{1,2} \and Liesbeth~M.C.~Janssen\inst{1,2}\thanks{E-mail: \email{l.m.c.janssen@tue.nl}} \and Simone~Ciarella\inst{1,3}\thanks{E-mail: \email{simoneciarella@gmail.com}} }
\institute{
  \inst{1} Department of Applied Physics, Eindhoven University of Technology -  P.O.~Box 513, NL-5600 MB Eindhoven, The Netherlands \\
  \inst{2} Institute for Complex Molecular Systems, Eindhoven University of Technology - P.O.~Box 513, NL-5600 MB Eindhoven, The Netherlands \\
  \inst{3} Laboratoire de Physique de l\textquotesingle Ecole Normale Sup\'erieure, ENS, Universit\'e PSL, CNRS, Sorbonne Universit\'e, Universit\'e de Paris - F-75005 Paris, France
}
% Only one author is allowed in the running head, the \shortauthor macro has to be used:
\shortauthor{Giulia Janzen \etal}

\date{\today}
\abstract{
A longstanding open question in the field of dense disordered matter is how precisely structure and dynamics are related to each other. With the advent of machine learning, it has become possible to agnostically predict the dynamic propensity of a particle in a dense liquid based on its local structural environment. Thus far, however, these machine-learning studies have focused almost exclusively on simple liquids composed of passive particles. Here we consider a mixture of both passive and active (i.e.\ self-propelled) Brownian particles, with the aim to identify the active particles from minimal local structural information. We compare a  state-of-the-art machine learning approach for passive systems with a new method we develop based on Voronoi tessellation. Both methods accurately identify the active particles based on their structural properties at high activity and low concentrations of active particles. Our Voronoi method is, however, substantially faster to train and deploy because it requires fewer, and easy to compute, input features. Notably, both become ineffective when the activity is low, suggesting a fundamentally different structural signature for dynamic propensity and non-equilibrium activity. Ultimately, these efforts might also find relevance in the context of biological active glasses such as confluent cell layers, where subtle changes in the microstructure can hint at pathological changes in cell dynamics. 
\vspace{-5mm}
}
\maketitle

\section{Introduction}

A central notion in the study of active particulate matter---systems of discrete entities which consume energy to perform work and move autonomously---is that the presence of activity can dramatically alter both spatial organization and (collective) dynamics~\cite{bechinger16}. 
The fact that active matter is intrinsically out-of-equilibrium renders the standard tools of statistical physics of limited use and leaves open the question of which quantifiers most accurately characterize and predict the dynamics of active matter~\cite{marchetti13}. This is a profound issue, especially in densely disordered phases such as liquids and glasses, where the relation between spatial structure and emergent dynamics is notoriously obscure~\cite{bechinger16,Janssen2019}.
A better understanding of structure-dynamics relations in active matter would be highly desirable both from a fundamental and more applied perspective. Notably, in the context of biological tissues and confluent cell layers, subtle structural changes can correlate with the motile properties of the individual cells, with relevance in processes such as cancer metastasis and embryonic development \cite{blauth21,Grosser21}. 

\begin{figure*}
    \centering
    \includegraphics[width=\textwidth]{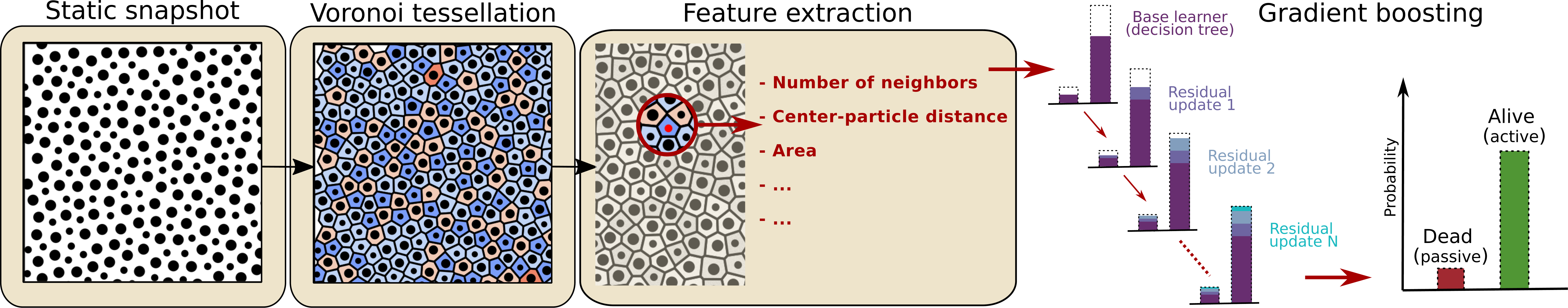}
    \vspace{-6mm}
    \caption{
    Sketch of our Voronoi-based machine learning approach for the identification of active particles in an active/passive mixture.
    The snapshot is processed by constructing its Voronoi tessellation, then extracting particle-specific features. A gradient boosting model is trained to evaluate the probability that each particle is dead (passive) or alive (active), only using the static features extracted from the single snapshot.
    }
    \label{fig:sketch}
\end{figure*}

Recent efforts have demonstrated that machine learning (ML) approaches are extremely effective in finding simple structural indicators that predict dynamical properties in densely disordered (near-)equilbrium systems~\cite{PhysRevLett.114.108001,doi:10.1126/science.aai8830,CubukJPC2016,Sussman2018,Boattini2018,Boattini2019,Schoenholz2016,Bapst2020,Paret2020,PhysRevE.101.010602,Boattini2021,Alkemade2022,oyama22,tah22,jung22,ciarella23mlmct,coslovich22,ciarella22tls,Alkemade23,janzen23ageing}. These findings firmly establish a correlation between local structure and the propensity of passive particles to move in a crowded environment.
While ML has also been applied with considerable success in purely active systems~\cite{cichos20,newby18,jeckel19,bo19,munozgil20,tah21,bag21,ruizgarcia22,janzen23ageing}, in practice and particularly in biological settings the system of interest will generally contain actors whose activity parameters are different, and distributed. In tumor tissue, for instance, there may be a distribution of epithelial (more stationary, passive) phenotypes and mesenchymal (more motile, active) ones, with %that are respectively identified as \textit{dead} or \textit{alive} elements. 
%Generically, 
the presence of mesenchymal cells being generally associated with greater metastatic potential~\cite{Grosser21,coban21}. 
Thus, developing the ability to reliably distinguish active %(`alive') 
from passive %(`dead') 
elements in dense collectives holds  diagnostic and prognostic potential. 

Inspired by this challenge, our work addresses a seemingly simple issue: Can we identify the active species in a binary model system of active and passive particles? Here we refer to passive entities as `dead' particles that do not possess any self-propulsion, while the active, self-propelled particles are `alive'. Naturally, the distinction would be most conveniently done based on dynamical (i.e.\ time-resolved) information, but in biological settings such dynamical data may be difficult to obtain. This is why we complicate the challenge considerably, and demand that the identification is performed based solely on static images (`snapshots') of the system. 

Recent studies~\cite{mejia2011bias,benichou2013biased,Wittkowski_2017,vasilyev2017cooperative,Banerjee2022} have shown that the presence of a small number of active particles within a dense passive bath can induce specific structural inhomogeneities. Explicitly, bath particles tend to accumulate at the front of an active particle while leaving a depletion zone behind it. This spatial anisotropy is usually irrelevant in the absence of activity, and is therefore also frequently ignored in ML methods designed for purely passive systems~\cite{Alkemade2022,jung22,coslovich22,zhang2023anisotropic}. However, it could be essential to distinguish dead particles from those that are alive.

In this work, we quantify the spatial anisotropy of each particle's local  environment through a simple Voronoi tesselation~\cite{zhang2023anisotropic,aurenhammer2000voronoi}; from this instantaneous structural information, we seek to predict the particle identity (dead or alive) using ML. We show that when the active particle fraction is low and the activity level is high, the shape of the Voronoi polygons around active particles exhibits distinct characteristics compared to those around passive particles, thus providing sufficient static information to reliably distinguish the two species. This observation highlights the effectiveness of Voronoi tessellation as a suitable tool for quantifying unique structural signatures associated with active particles, at least within a suitably chosen parameter range.

We also compare our method with an existing state-of-the-art ML approach based on particle configurations, originally developed for purely passive simulated supercooled liquids~\cite{Boattini2021,Alkemade2022,coslovich22}. 
We demonstrate that both approaches are only successful when the fraction of active particles is low and the activity is high. 
Furthermore, we show that the Voronoi method is substantially faster and easier to compute because it requires fewer and cheaper input features. 
This emphasizes the significance of anisotropy in telling apart dead from alive. 
However, when the active particle concentration is not dilute or the active force is too small, all these static approaches fail to provide accurate predictions, suggesting that the instantaneous local structure no longer fully encodes the particle identity. For this more challenging regime, we develop a pseudo-static approach, detailed in the supplementary material (SM), which also incorporates information on the averaged statistical \textit{fluctuations} of the local structure.

Briefly, our Voronoi approach to distinguish between active and passive particles from a single snapshot is outlined in Fig.~\ref{fig:sketch}. First, the snapshots are processed by performing a Voronoi tessellation~\cite{aurenhammer2000voronoi}, from which we compute $11$ relevant features 
that effectively capture local anisotropies.
At the core of our Voronoi approach is a gradient boosting model, discussed in the methods section, that takes into account the local environment of each particle to predict if it is dead or alive.
Furthermore, the predictions generated by our ML model are explainable, enhancing the interpretability of the results.

Ultimately, our findings lead us to conclude that there are two distinct regions within the parameter space defined by the fraction of active particles and the active force. In the first region, characterized by high activity and a low fraction of active particles, active and passive particles can be distinguished effectively using a static approach. However, in the second region, active and passive particles become indistinguishable from a static point of view.

%%%%%%%%%%%%%%%%%%%%%%%%%%%%%%%%%%%%%%%%%%%%%%%%%%%%%%%5%------------

\section{Methods}\label{sec:methods}
\subsection{Simulation model}
\noindent Our model system is a two-dimensional (2D) Kob-Andersen mixture~\cite{Kob1994,Michele2004,Flenner2005} composed of 650 and 350 particles of type A and B, respectively. We extend the model by turning a fraction $\phi_a$ into active Brownian particles~\cite{Romanczuk2012active,Ramaswamy2017active,Lowen2020active,Hagen2011ABP} (same proportion for each particle type). We have also conducted similar investigations in a three-dimensional (3D) mixture, comprising 800 particles of type A and 200 particles of type B, and obtained similar results.

The self-propelled motion is characterized by a constant active force of magnitude $F_a$. 
To develop a model that can be applied to study highly crowded biological systems, such as dense cellular collectives, we have chosen to focus on steady-state configurations with a density of $\rho=1.2$ and temperature of $T=0.5$ (in simulation units) and retrieve 2000 different independent configurations for each studied setting. Moreover, to effectively characterize the large activity regime, we consider activities ranging from 0.5 to 100, as cells primarily move due to active motion. Furthermore, we have verified that similar results are obtained at a temperature of $T=0.6$. Finally, to confirm that the dynamics at high activities are not influenced by finite-size effects, we performed the same analysis on a larger system with 10\,000 particles, and we obtained similar results. All the details of the simulation model and the data collection are reported in the SM.

\subsection{Gradient Boosting model}
\label{sec:gb}
\noindent We treat the identification of the active particles as a binary classification problem. For this we use LightGBM, which is a gradient-boosting machine-learning method based on decision trees~\cite{ke2017lightgbm}. This algorithm is widely used due to its efficiency, accuracy, and interpretability, and has been successfully employed e.g.\ in the context of low-temperature glassy materials~\cite{ciarella22tls}. We show in the SM that it outperforms neural networks for our task of  distinguishing active from passive particles. 

In general, gradient boosting decision trees~\cite{friedman01} comprise an ensemble learning method that combines the power of decision trees~\cite{kotsiantis13} and gradient boosting. It involves building a sequence of decision trees, where each subsequent tree corrects the errors made by the previous ones, by performing a residual update step. By iteratively adding decision trees and adjusting their weights based on the residuals, the model gradually improves its ability to make accurate predictions. 

LightGBM introduces Gradient-based One-Side Sampling and
Exclusive Feature Bundling~\cite{ke2017lightgbm} in order to avoid the need to scan all the data to update the residuals at each step. As a result, gradient boosting can be used for large datasets, such as the collection of simulation snapshots used in this work. 
In the end, after calculating all the relevant structural input features (discussed in the next section), the training of our model takes only several minutes on a standard laptop. 

To quantify our ML performance we use the accuracy, defined as the number of correct predictions divided by the total number of predictions. 
When the fraction of active particles $\phi_a\neq 0.5$ we simply discard a subset of the particles from the data such that the number of active and passive species is equal. This allows us to work with a balanced dataset, which enhances the classifier's generalizability.
Since our input features are particle-resolved and additional data collection is quite inexpensive, this presents no practical challenges.

\subsection{Structural input features}

Our main goal is to distinguish the active particles from the passive ones using only instantaneous static information, i.e.\ a single snapshot. 
For this, we employ two alternative sets of local, particle-resolved structural properties as input for our machine-learning model. We refer to these as (i) the \textit{shell-based approach}, developed earlier for purely passive systems \cite{Boattini2021}, and (ii) the \textit{Voronoi approach}, which we introduce here  to capitalize on the information contained in local anisotropy.

The features of the \textit{shell-based approach} have been introduced to effectively predict the dynamic propensity of purely passive particles~\cite{Boattini2021,Alkemade2022}.
Since the dynamic propensity quantifies the average squared displacement of a particle from its initial condition~\cite{berthier07,widmercooper07}, and active particles have an additional self-propelling force, it is natural to expect that active particles have a larger dynamic propensity.
We also confirm this observation by measuring the mean-squared displacement~\cite{PhysRevE.96.062608,RevModPhys.88.045006}, reported in the SM. Notice that similar indicators also predict localized plastic events~\cite{Richard2020,ridout22}.
This means that if the dynamical information remains encoded in the structure even for active systems, then this approach should be able to pick it up and identify the active particles from their enhanced propensity. 

To implement the shell-based approach we follow the definitions of Ref.~\cite{Boattini2021}. Briefly, these features consist of $n$th-order radial and angular descriptors, denoted as $G_i^{(n)}$ and $q_i^{(n)}$, respectively. The former measures for each particle $i$ the local particle density within a radial shell, while the latter expands the local density within a shell in terms of spherical harmonics; the order $n$ indicates the degree of averaging over the structural features of nearby particles. Here we consider $n=0, 1, 2$. For $n=0$, we compute $50$ radial descriptors per species with radial distance $r \in [0,5]$ and shell width $\delta=0.1$, and $192$ angular descriptors with $r \in [1,2.5]$, $\delta=0.1$, and spherical harmonics of order $l \in [1,12]$. In sum, we use a total of 876 static quantities as input in the shell-based approach.

As an alternative to the shell-based features of Ref.~\cite{Boattini2021}, we introduce the \textit{Voronoi approach}, which naturally quantifies the local anisotropy created by the active particles. Briefly, we perform a Voronoi tessellation of all particle positions and compute the features based on the shape of the Voronoi polygon around each particle. Explicitly, we consider the number of neighbors, the polygon area and perimeter, the distance between the center of the polygon and the actual particle position, the maximum and minimum distances between nearest neighbors, the maximum and minimum distances between the vertices of a polygon, the maximum and minimum distances between the particle position and its polygon vertices, and finally we add the particle type (A or B). Overall this amounts to only 11 relatively simple input features.
Compared to the shell-based approach, the Voronoi approach is thus significantly cheaper.

\section{Results and Discussion}
\begin{figure}
    \centering
    \includegraphics[width=0.8\columnwidth]{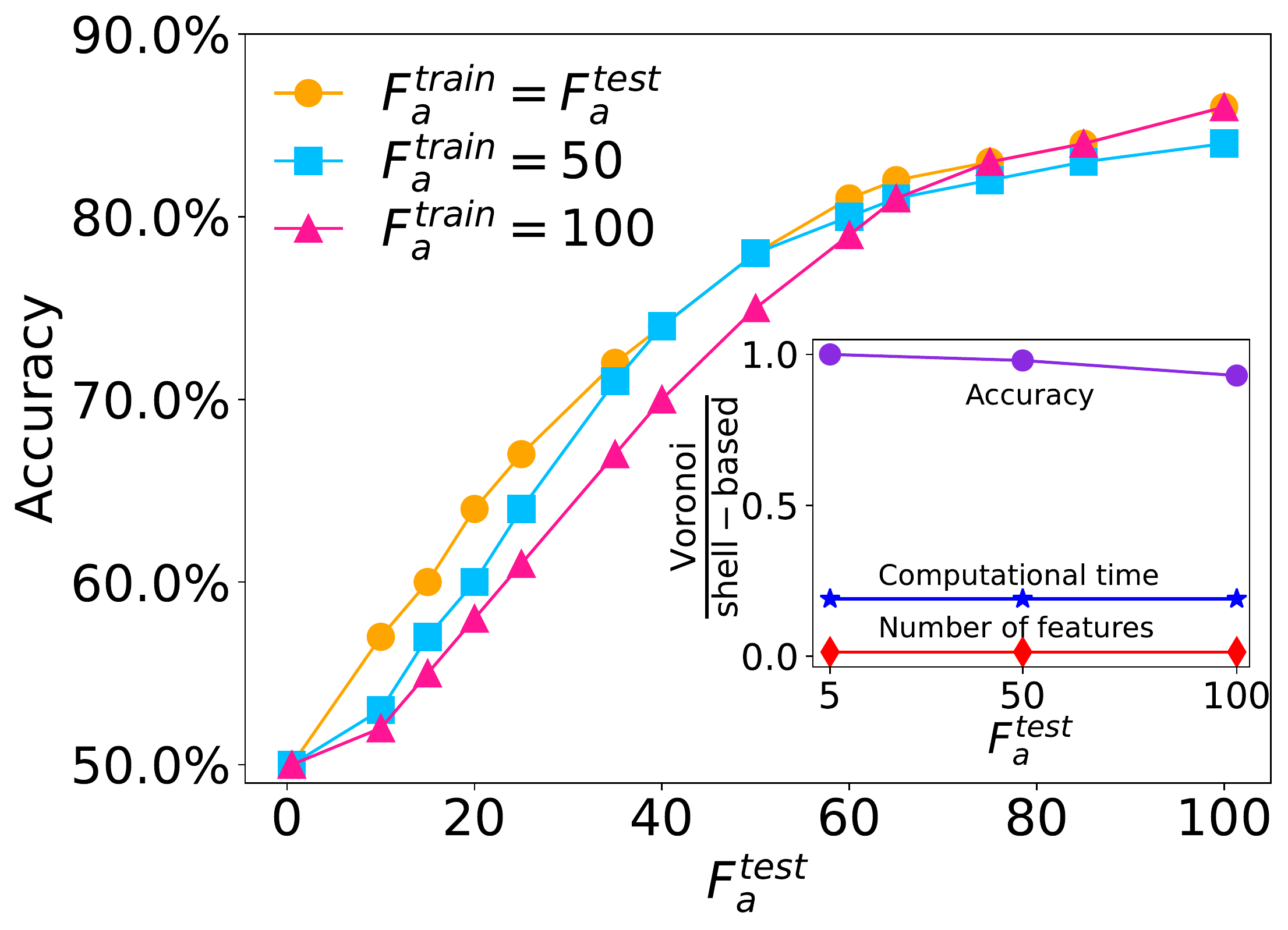}
    \vspace{-3mm}
    \caption{Accuracy of the Voronoi ML approach as a function of the active force $F_a=F_a^{\mathrm{test}}$, with $\phi_a=0.008$. The orange circles represent scores obtained from separate models, where each one was trained using $F_a^{\mathrm{train}}=F_a^{\mathrm{test}}$. The blue squares and pink triangles represent scores obtained from a single global model trained with data only for $F_a=50$ or $F_a=100$, respectively. The inset shows the comparison between the Voronoi and the shell-based approaches for three different active forces $F_{a}^{\mathrm{test}}=F_{a}^{\mathrm{train}}=5,50,100$. The purple circles, blue stars, and red diamonds represent the ratio between accuracy, computational times, and number of features, respectively.
    }
    \label{fig:model_acc_Fa}
\end{figure}
\begin{figure}
    \centering
    \includegraphics[width=0.8\columnwidth]{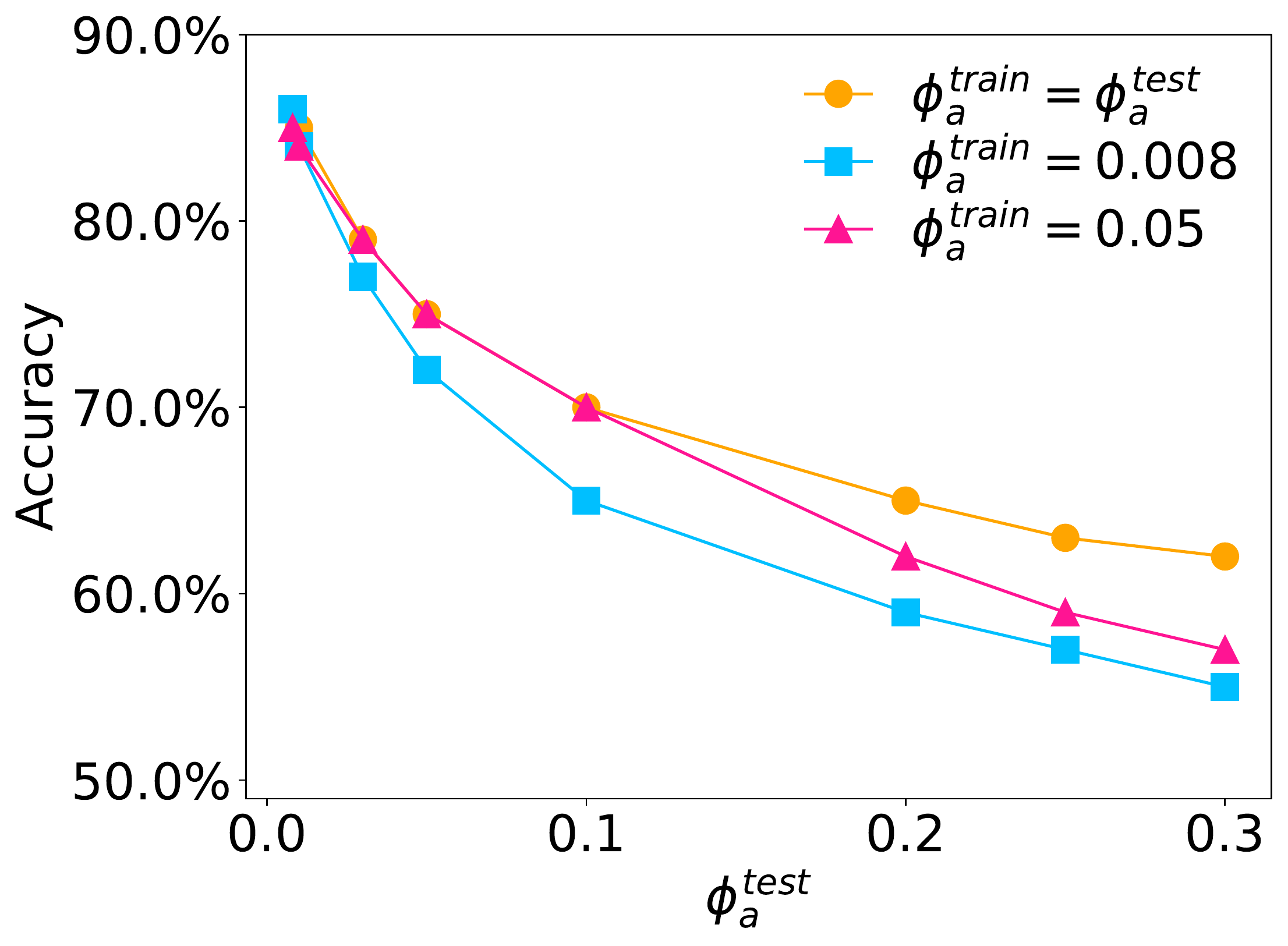}
    \vspace{-3mm}
    \caption{Accuracy of the Voronoi ML approach as a function of the fraction of active particles $\phi_a=\phi_a^{\mathrm{test}}$, with $F_a=100$. The orange circles represent scores obtained from separate models, where each one was trained using $\phi_a^{\mathrm{train}}=\phi_a^{\mathrm{test}}$. The blue squares and pink triangles represent scores obtained from a single global model trained with data only for $\phi_a=0.008$ or $\phi_a=0.05$, respectively.
    }
    \label{fig:model_acc_phia}
\end{figure}
\begin{figure}
    \centering
    \includegraphics[width=0.9\columnwidth]{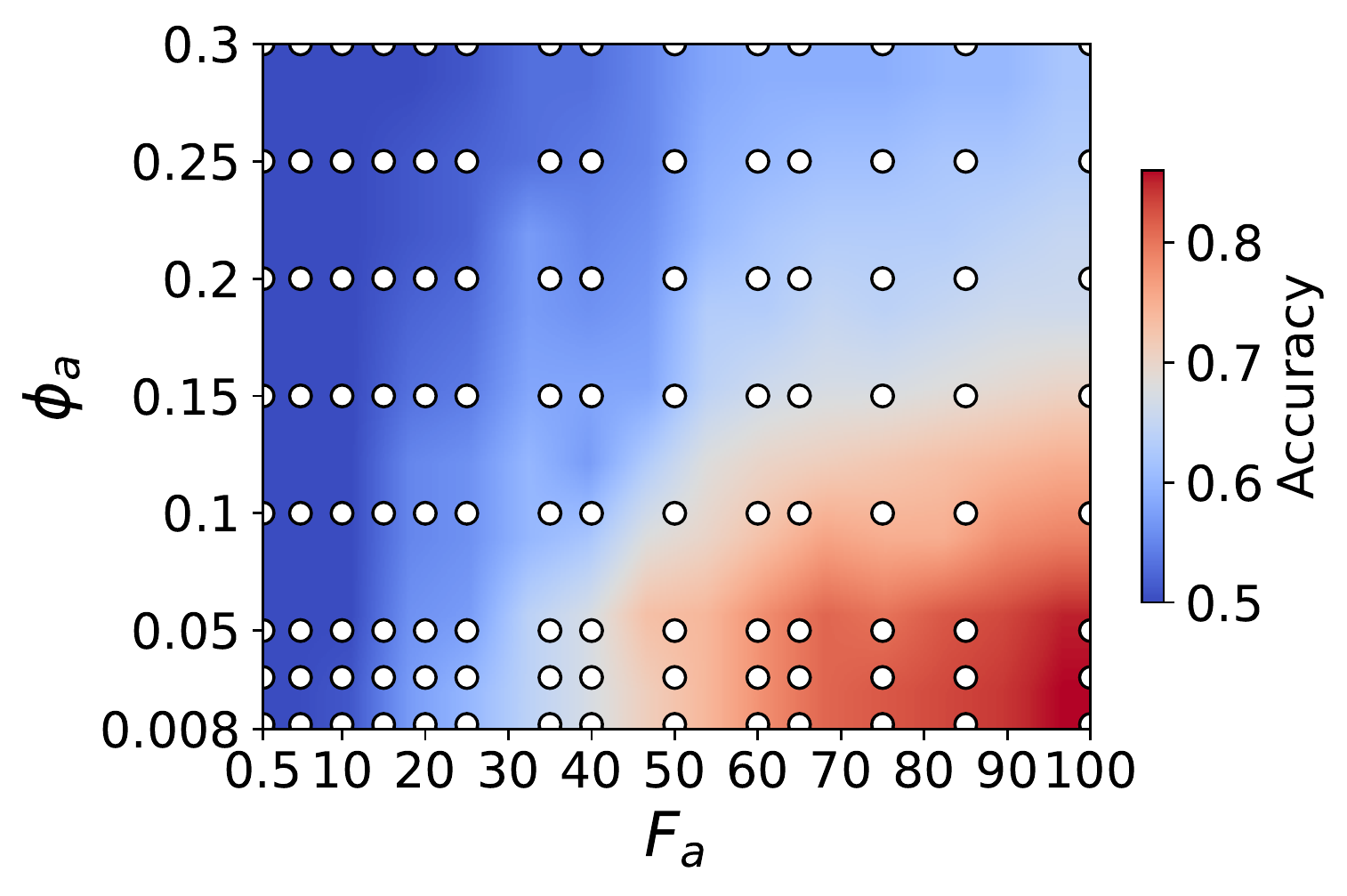}
    \vspace{-4mm}
    \caption{Accuracy map of the Voronoi ML approach in the ($F_a, \phi_a$) plane for a 2D system at temperature $T=0.5$ and density $\rho=1.2$. %and $\tau_r=1$.
    }
    \label{fig:heatmap}
\end{figure}
%%%%%%%%%%%%%%%%%%%%%%%%%%%%%%%%%%%%%%%%
\subsection{Distinguishing dead from alive}
We report in Fig.~\ref{fig:model_acc_Fa} the accuracy achieved by our Voronoi ML model in the classification of an active/passive mixture with $\phi_a=0.008$, for different values of the active force $F_a$. We compare the accuracy that we get when the model is trained and tested at a single specific value of $F_a$ (orange circles), with the accuracy of a model trained when fixing $F_a=50$ (blue squares) or $F_a=100$ (pink triangles). All three curves produce rather similar accuracies, implying that a single ML model trained at a fixed $F_a$ can also produce good predictions for unseen parameter regimes. This shows reasonable generalizability of the model. However, the accuracy in all three cases drops significantly for lower activity. We attribute this to the fact that small values of $F_a$ render it extremely challenging to differentiate between structural signatures induced by either active forces or passive Brownian motion. Hence, any static approach based purely on instantaneous structural properties fails.

The inset of Figure \ref{fig:model_acc_Fa} demonstrates that very similar accuracies can be achieved when training the ML algorithm with the more intricate structural properties  $G_i^{(n)}$ and $q_i^{(n)}$ as input (shell-based approach).  
However, the shell-based approach is considerably more computationally expensive compared to the Voronoi method. The reason is that the shell-based approach employs almost two orders of magnitude more features. Consequently, we can conclude that, for this specific classification task, the Voronoi approach is more efficient. Moreover, we find similar results for both the Voronoi and the shell-based approach in 3D (see SM). 

Figure~\ref{fig:model_acc_phia} reports the performance of the Voronoi active/passive classifiers as a function of the percentage of active particles $\phi_a$, while keeping $F_a$ fixed at 100.
For $\phi_a \leq 0.1$, our approach demonstrates excellent predictive capability when the ML model is trained and tested at a specific $\phi_a$ value (represented by orange circles). Additionally, we compare the accuracy obtained when training the model with $\phi_a$ fixed at 0.008 (blue squares) or 0.05 (pink triangles). 
The latter case yields better transferability than when training only on $\phi_a=0.008$, especially in the region $\phi_a\leq 0.1$, where it rivals the performance of the model trained at each $\phi_a$ value separately. 
Noticeably, however, 
when the fraction of active particles $\phi_a$ becomes large, the accuracy drops to below 70\%, even for our best ML model. This indicates that the instantaneous structural input becomes less informative to differentiate between active and passive particles, which we attribute to the fact that the presence of more active particles increasingly affects the entire structure of the material and diminishes the front-wake asymmetry.
Finally, we have also repeated the Voronoi tessellation analysis at a lower density of $\rho=1.0$ with a fraction of active particles $\phi_a=0.008$. Interestingly, the results are qualitatively similar to those obtained at $\rho=1.2$, but with a more pronounced anisotropy observed at a lower activity ($F_a=50$) compared to the higher-density system (see SM for more details).

A full overview of the Voronoi ML model accuracy in the $(F_a, \phi_a)$ plane is reported in Fig.~\ref{fig:heatmap}. 
This figure reveals two distinct regions: a red region representing a `distinguishable' area and a blue region representing an `indistinguishable' area.  In the red region, characterized by low $\phi_a$ and high $F_a$ values, the algorithm achieves an accuracy greater than 0.75, enabling it to differentiate adequately between active and passive particles based on a single snapshot. In this region, the structure surrounding the active particle significantly differs from that of the passive particle. As a result, even a shell-based approach, which does not consider the directional dependence of passive particle density around the active particle, still yields accurate results (inset of Fig.~\ref{fig:model_acc_Fa}). 

In the blue region, characterized by either high $\phi_a$ or low $F_a$, any static approach (either shell-based or Voronoi) is insufficient to effectively differentiate between active and passive particles. When $F_a<40$, the activity is not high enough to induce noticeable spatial inhomogeneities. On the other hand, when $\phi_a$ is too high, the interaction among active particles complicates the classification task. Particularly, in cases where both $F_a$ and $\phi_a$ are high, the system can undergo motility-induced phase separation (MIPS) \cite{Wittkowski_2017}. %PhysRevLett.114.018301,cates2015motility,Wittkowski_2017,C9SM01803D}. 
In view of the poor overall performance of our ML method for both high $\phi_a$ and low $F_a$, we conclude that if a local structure-dynamics relationship exists in active systems, it must be significantly different from that observed in passive systems, as indicated by the failure of the shell-based approach. Additionally, this relationship is not hidden in the anisotropy, as indicated by the failure of the Voronoi approach.

In order to still achieve good predictability in the regime where a purely static ML approach fails (blue region in Fig.~\ref{fig:heatmap}), we have developed a so-called pseudo-static method. 
This approach, detailed in the SM, is able to identify active particles at the cost of more input information.
Briefly, while not requiring a fully time-resolved dynamical trajectory, it requires some knowledge of the averaged statistical \textit{fluctuations} of the local structure. It is possible to obtain these statistics from a collection of snapshots that do not have to be time-ordered, but in which each particle has to be tracked. 
This tracking information makes the method naturally more computationally expensive.
Overall, in the pseudo-static approach, the information about the local structure is complemented by statistical measurements of the same features at different times. We have verified that using this information on the structural fluctuations, an accuracy of 95\%-100\% can be achieved (see SM). 
\begin{figure}
    \centering
    \includegraphics[width=0.8\columnwidth]{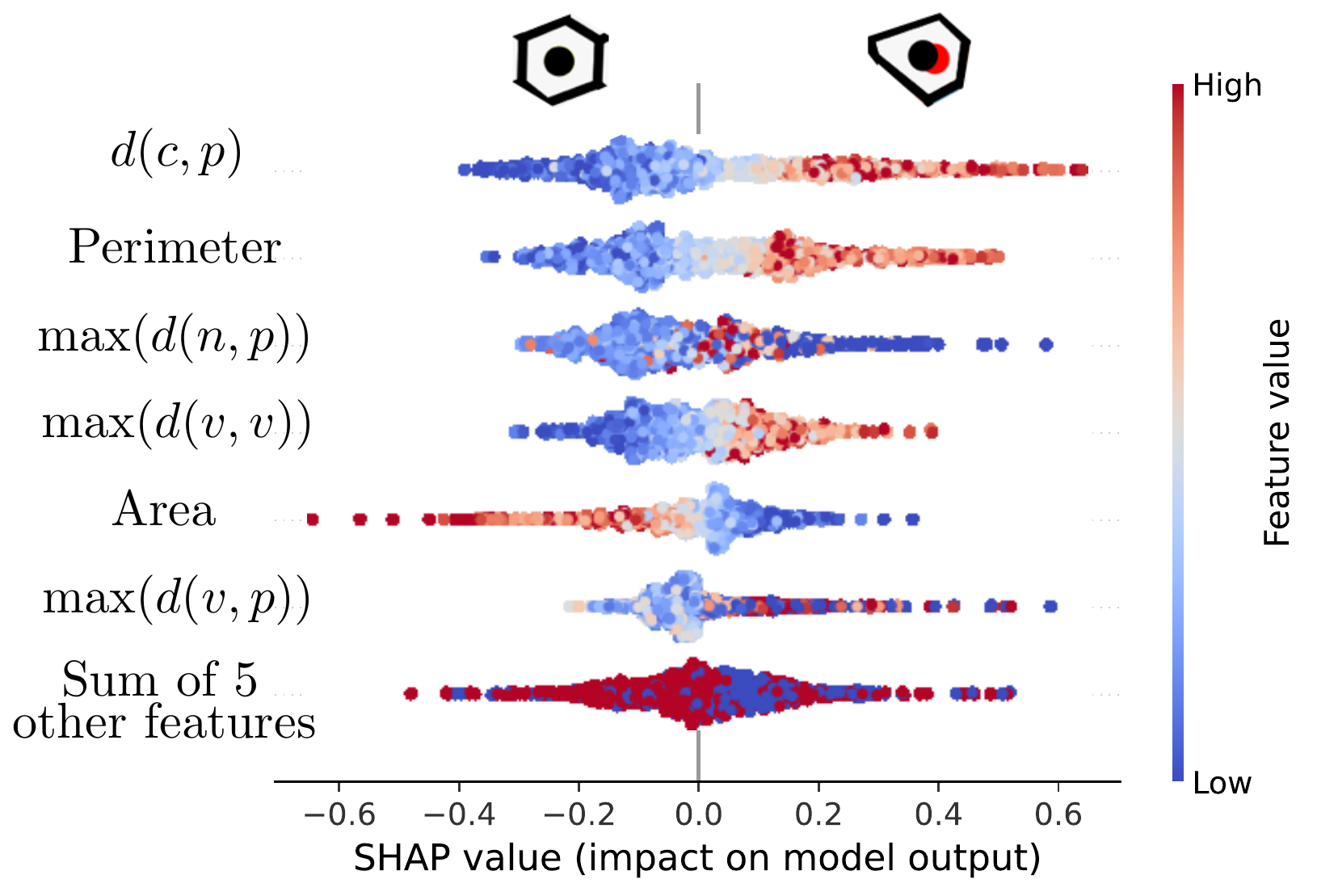}
    \caption{The six most important features for the Voronoi model, represented as a SHAP beeswarm plot at $F_a=100$ and $\phi_a=0.008$. The position of the dots is determined by the SHAP values of the features, and the color is used to display the value of the features. The top two polygons in the figure are a detail of a snapshot that represents the typical shapes observed in the passive (left) and active (right) cases. In the passive scenario, the center of the polygon aligns with the particle position, whereas in the active case, they do not coincide.
    }
    \label{fig:shapFa100}
\end{figure}

\subsection{Model explanation}
Let us return to the parameter regime where our Voronoi ML model is most successful in distinguishing active from passive particles, i.e.\ small $\phi_a$ and large $F_a$. In this regime we can interpret the decisions made by the ML model. To do so, we calculate the model explanations using SHapley Additive exPlanation (SHAP) analysis \cite{lundberg2017unified}. Briefly, this method, based on cooperative game theory, calculates Shapley values that distribute the model's prediction among the features while considering their individual contributions to the output. We apply this analysis to the Voronoi model trained on $F_a=100$ and $\phi_a=0.008$.

Figure~\ref{fig:shapFa100} shows the SHAP beeswarm plot, which indicates the six most important features and how the values of these features influence the model's predictions. The first six features are, in order of importance: the distance between the center of the polygon and the particle [$d(c,p)$], the perimeter of the polygon, the maximum distance between the particle and its neighbors [$\mathrm{max}(d(n,p))$], the maximum distance between the vertices of the polygon [$\mathrm{max}(d(v,v))$], the area of the polygon, and the maximum distance between the particle and the vertices of the polygon [$\mathrm{max}(d(v,p))$]. 
The colors show that the model interprets low values of $d(c,p)$ as a passive particle and high values of $d(c,p)$ as an active particle. Considering that passive particles tend to accumulate in front of active particles while creating a void behind them, we anticipate the polygons around active particles to exhibit elongation. Consequently, due to this anisotropy, we expect the active particle to be situated at a different position compared to the center of the polygon, as we confirm in representative polygons shown at the top of Fig.~\ref{fig:shapFa100}. 
The SHAP analysis clearly validates this expectation. Hence, rather than using the ML model solely as a black box, we infer that the model captures the correct physical picture and confirms the importance of spatial anisotropy induced by activity.  
Lastly, the SHAP analysis has also been performed in 3D, and the corresponding results can be found in the SM.

\section{Conclusions}
The focus of this work is to identify active (alive) entities in a densely disordered mixture of active and passive (dead) particles, using only a single picture of the system as input. 
We compare two static ML approaches: one based on the Voronoi tessellation and another employing a shell-based approach that utilizes descriptors previously employed in passive systems to predict dynamic particle propensities~\cite{Sussman2018,Boattini2018,Boattini2019,Bapst2020,Paret2020,Boattini2021}.

We find that a static approach based on a single snapshot is able to predict whether a particle is dead or alive when the activity is high and the fraction of active particles is low, in both 2D and 3D. However, this approach fails when the activity is low or the fraction of active particles is high, implying a different 
structural signature for activity and dynamic propensity in passive systems. Moreover, we observe that while the Voronoi and shell-based approaches yield similar accuracies, the shell-based approach is computationally more expensive.

Within the region of the $(F_a,\phi_a)$ plane where the static approach is effective for particle classification, our model exhibits reasonably good performance when extrapolating to active forces and compositions outside its training range, demonstrating its robustness. This transferability can be particularly useful in experiments, where the active force or the precise number of active species may be difficult to quantify.

Additionally, we explain the decisions of our ML model in order to understand how it is able to identify the active entities from the Voronoi input. We find that the distance between a particle and the center of its corresponding polygon is the most significant feature for this classification task, serving as a simple and reliable measure of the distinct spatial anisotropy surrounding an active particle.

We hypothesize that the region of the $(F_a,\phi_a)$ plane where a purely static approach fails depends on system parameters such as density and temperature. Indeed, when considering a lower density, we have found that our classifier can perform better for certain levels of activity (SM). However, while the boundary of the distinguishable region may thus be varied with external control parameters,
we still expect a region where active and passive particles remain indistinguishable from a single snapshot. 
 In this regime more information is inevitably needed, as confirmed by our pseudo-static approach that incorporates additional knowledge of structural \textit{fluctuations}.

In summary, this work demonstrates that the shape of the polygon is sufficient to differentiate between active and passive particles in a specific region of the $(F_a,\phi_a)$ plane. This work serves as a step to better understand the elusive structure-dynamics relation in densely disordered non-equilibrium systems. We believe that our approach can also be a valuable tool for investigating experimental systems such as biological cells, where identifying the most active entities visually can be challenging. It is thus our hope that the here presented approach can help to process large experimental datasets and contribute to the discovery of new connections between structural and dynamical properties in complex non-equilibrium materials. 

\begin{acknowledgements}
This work has been financially supported by the Dutch Research Council (NWO) through a START-UP grant (VED, CL, and LMCJ), Physics Projectruimte grant (GJ and LMCJ), ENW-XL grant (CS and LMCJ), and Vidi grant (LMCJ).
\end{acknowledgements}

\bibliographystyle{eplbib}
\bibliography{references,library,abp}

\begin{thebibliography}{10}
\expandafter\ifx\csname url\endcsname\relax\def\url#1{\texttt{#1}}\fi

\bibitem{bechinger16}
\Name{Bechinger C., Leonardo R.~D., L\"owen H., Reichhardt C., Volpe G. \and
  Volpe G.} \REVIEW{Reviews of Modern Physics}{88}{2016}{45006}.

\bibitem{marchetti13}
\Name{Marchetti M.~C., Joanny J.~F., Ramaswamy S., Liverpool T.~B., Prost J.,
  Rao M. \and Simha R.~A.} \REVIEW{Reviews of Modern Physics}{85}{2013}{1143}.

\bibitem{Janssen2019}
\Name{Janssen L. M.~C.} \REVIEW{Journal of Physics: Condensed
  Matter}{31}{2019}{503002}.

\bibitem{blauth21}
\Name{Blauth E., Kubitschke H., Gottheil P., Grosser S. \and K\"as J.~A.}
  \REVIEW{Frontiers in Physics}{9}{2021}{}.

\bibitem{Grosser21}
\Name{Grosser S., Lippoldt J., Oswald L., Merkel M., Sussman D.~M., Renner F.,
  Gottheil P., Morawetz E.~W., Fuhs T., Xie X., Pawlizak S., Fritsch A.~W.,
  Wolf B., Horn L.-C., Briest S., Aktas B., Manning M.~L. \and K\"as J.~A.}
  \REVIEW{Physical Review X}{11}{2021}{11033}.

\bibitem{PhysRevLett.114.108001}
\Name{Cubuk E.~D., Schoenholz S.~S., Rieser J.~M., Malone B.~D., Rottler J.,
  Durian D.~J., Kaxiras E. \and Liu A.~J.} \REVIEW{Phys. Rev.
  Lett.}{114}{2015}{108001}.

\bibitem{doi:10.1126/science.aai8830}
\Name{Cubuk E.~D., Ivancic R. J.~S., Schoenholz S.~S., Strickland D.~J., Basu
  A., Davidson Z.~S., Fontaine J., Hor J.~L., Huang Y.-R., Jiang Y., Keim
  N.~C., Koshigan K.~D., Lefever J.~A., Liu T., Ma X.-G., Magagnosc D.~J.,
  Morrow E., Ortiz C.~P., Rieser J.~M., Shavit A., Still T., Xu Y., Zhang Y.,
  Nordstrom K.~N., Arratia P.~E., Carpick R.~W., Durian D.~J., Fakhraai Z.,
  Jerolmack D.~J., Lee D., Li J., Riggleman R., Turner K.~T., Yodh A.~G.,
  Gianola D.~S. \and Liu A.~J.} \REVIEW{Science}{358}{2017}{1033}.

\bibitem{CubukJPC2016}
\Name{Cubuk E.~D., Schoenholz S.~S., Kaxiras E. \and Liu A.~J.} \REVIEW{The
  Journal of Physical Chemistry B}{120}{2016}{6139} pMID: 27092716.

\bibitem{Sussman2018}
\Name{Sussman D.~M., Paoluzzi M., Marchetti M.~C. \and Manning M.~L.}
  \REVIEW{EPL (Europhysics Letters)}{121}{2018}{36001}.

\bibitem{Boattini2018}
\Name{Boattini E., Ram M., Smallenburg F. \and Filion L.} \REVIEW{Molecular
  Physics}{116}{2018}{3066}.

\bibitem{Boattini2019}
\Name{Boattini E., Dijkstra M. \and Filion L.} \REVIEW{The Journal of Chemical
  Physics}{151}{2019}{154901}.

\bibitem{Schoenholz2016}
\Name{Schoenholz S.~S., Cubuk E.~D., Sussman D.~M., Kaxiras E. \and Liu A.~J.}
  \REVIEW{Nat. Phys.}{12}{2016}{469}.

\bibitem{Bapst2020}
\Name{Bapst V., Keck T., Grabska-Barwi{\'{n}}ska A., Donner C., Cubuk E.~D.,
  Schoenholz S.~S., Obika A., Nelson A. W.~R., Back T., Hassabis D. \and Kohli
  P.} \REVIEW{Nature Physics}{16}{2020}{448}.

\bibitem{Paret2020}
\Name{Paret J., Jack R.~L. \and Coslovich D.} \REVIEW{The Journal of Chemical
  Physics}{152}{2020}{144502}.

\bibitem{PhysRevE.101.010602}
\Name{Landes F. m. c.~P., Biroli G., Dauchot O., Liu A.~J. \and Reichman D.~R.}
  \REVIEW{Phys. Rev. E}{101}{2020}{010602}.

\bibitem{Boattini2021}
\Name{Boattini E., Smallenburg F. \and Filion L.} \REVIEW{Physical Review
  Letters}{127}{2021}{88007}.

\bibitem{Alkemade2022}
\Name{Alkemade R.~M., Boattini E., Filion L. \and Smallenburg F.} \REVIEW{The
  Journal of Chemical Physics}{156}{2022}{204503}.

\bibitem{oyama22}
\Name{Oyama N., Koyama S. \and Kawasaki T.} \REVIEW{arXiv preprint
  arXiv:2208.00349}{}{2022}{}.

\bibitem{tah22}
\Name{Tah I., Ridout S.~A. \and Liu A.~J.} \REVIEW{The Journal of Chemical
  Physics}{157}{2022}{124501}.

\bibitem{jung22}
\Name{Jung G., Biroli G. \and Berthier L.} \REVIEW{arXiv preprint
  arXiv:2210.16623}{}{2022}{}.

\bibitem{ciarella23mlmct}
\Name{Ciarella S., Chiappini M., Boattini E., Dijkstra M. \and Janssen L.
  M.~C.} \REVIEW{Machine Learning: Science and Technology}{4}{2023}{025010}.

\bibitem{coslovich22}
\Name{Coslovich D., Jack R.~L. \and Paret J.} \REVIEW{The Journal of Chemical
  Physics}{157}{2022}{204503}.

\bibitem{ciarella22tls}
\Name{Ciarella S., Khomenko D., Berthier L., Mocanu F.~C., Reichman D.~R.,
  Scalliet C. \and Zamponi F.} \REVIEW{arXiv preprint
  arXiv:2212.05582}{}{2022}{}.

\bibitem{Alkemade23}
\Name{Alkemade R.~M., Smallenburg F. \and Filion L.} \REVIEW{arXiv preprint
  arXiv:2301.13106}{}{2023}{}.

\bibitem{janzen23ageing}
\Name{Janzen G., Smit C., Visbeek S., Debets V.~E., Luo C., Storm C., Ciarella
  S. \and Janssen L. M.~C.} \REVIEW{arXiv preprint arXiv:2303.00636}{}{2023}{}.

\bibitem{cichos20}
\Name{Cichos F., Gustavsson K., Mehlig B. \and Volpe G.} \REVIEW{Nature Machine
  Intelligence}{2}{2020}{94}.

\bibitem{newby18}
\Name{Newby J.~M., Schaefer A.~M., Lee P.~T., Forest M.~G. \and Lai S.~K.}
  \REVIEW{Proceedings of the National Academy of Sciences}{115}{2018}{9026}.

\bibitem{jeckel19}
\Name{Jeckel H., Jelli E., Hartmann R., Singh P.~K., Mok R., Totz J.~F.,
  Vidakovic L., Eckhardt B., Dunkel J. \and Drescher K.} \REVIEW{Proceedings of
  the National Academy of Sciences}{116}{2019}{1489}.

\bibitem{bo19}
\Name{Bo S., Schmidt F., Eichhorn R. \and Volpe G.} \REVIEW{Physical Review
  E}{100}{2019}{10102}.

\bibitem{munozgil20}
\Name{{n}oz Gil G.~M., Garcia-March M.~A., Manzo C., Mart\'in-Guerrero J.~D.
  \and Lewenstein M.} \REVIEW{New Journal of Physics}{22}{2020}{013010}.

\bibitem{tah21}
\Name{Tah I., Sharp T.~A., Liu A.~J. \and Sussman D.~M.} \REVIEW{Soft
  Matter}{17}{2021}{10242}.

\bibitem{bag21}
\Name{Bag S. \and Mandal R.} \REVIEW{Soft Matter}{17}{2021}{8322}.

\bibitem{ruizgarcia22}
\Name{Ruiz-Garcia M., Gutierrez C. M.~B., Alexander L.~C., Aarts D. G. A.~L.,
  Ghiringhelli L. \and Valeriani C.} \REVIEW{arXiv preprint
  arXiv:2203.14846}{}{2022}{}.

\bibitem{coban21}
\Name{Coban B., Bergonzini C., Zweemer A. J.~M. \and Danen E. H.~J.}
  \REVIEW{British Journal of Cancer}{124}{2021}{49}.

\bibitem{mejia2011bias}
\Name{Mej{\'\i}a-Monasterio C. \and Oshanin G.} \REVIEW{Soft
  Matter}{7}{2011}{993}.

\bibitem{benichou2013biased}
\Name{B{\'e}nichou O., Illien P., Mejia-Monasterio C. \and Oshanin G.}
  \REVIEW{Journal of Statistical Mechanics: Theory and
  Experiment}{2013}{2013}{P05008}.

\bibitem{Wittkowski_2017}
\Name{Wittkowski R., Stenhammar J. \and Cates M.~E.} \REVIEW{New Journal of
  Physics}{19}{2017}{105003}.
\newline\url{https://dx.doi.org/10.1088/1367-2630/aa8195}

\bibitem{vasilyev2017cooperative}
\Name{Vasilyev O.~A., B{\'e}nichou O., Mej{\'\i}a-Monasterio C., Weeks E.~R.
  \and Oshanin G.} \REVIEW{Soft Matter}{13}{2017}{7617}.

\bibitem{Banerjee2022}
\Name{Banerjee J.~P., Mandal R., Banerjee D.~S., Thutupalli S. \and Rao M.}
  \REVIEW{Nat. Commun.}{13}{2022}{4533}.
\newline\url{https://doi.org/10.1038/s41467-022-31984-z}

\bibitem{zhang2023anisotropic}
\Name{Zhang H., Liu F. \and Han Y.} \REVIEW{arXiv preprint
  arXiv:2305.04179}{}{2023}{}.

\bibitem{aurenhammer2000voronoi}
\Name{Aurenhammer F. \and Klein R.} \REVIEW{Handbook of computational
  geometry}{5}{2000}{201}.

\bibitem{Kob1994}
\Name{Kob W. \and Andersen H.~C.} \REVIEW{Phys. Rev. Lett.}{73}{1994}{1376}.

\bibitem{Michele2004}
\Name{Michele C.~D., Sciortino F. \and Coniglio A.} \REVIEW{Journal of Physics:
  Condensed Matter}{16}{2004}{L489}.

\bibitem{Flenner2005}
\Name{Flenner E. \and Szamel G.} \REVIEW{Phys. Rev. E}{72}{2005}{1}.

\bibitem{Romanczuk2012active}
\Name{Romanczuk P., B{\"a}r M., Ebeling W., Lindner B. \and Schimansky-Geier
  L.} \REVIEW{The European Physical Journal Special Topics}{202}{2012}{1}.

\bibitem{Ramaswamy2017active}
\Name{Ramaswamy S.} \REVIEW{Journal of Statistical Mechanics: Theory and
  Experiment}{2017}{2017}{054002}.

\bibitem{Lowen2020active}
\Name{L{\"o}wen H.} \REVIEW{The Journal of Chemical
  Physics}{152}{2020}{040901}.

\bibitem{Hagen2011ABP}
\Name{ten Hagen B., van Teeffelen S. \and Löwen H.} \REVIEW{Journal of
  Physics: Condensed Matter}{23}{2011}{194119}.

\bibitem{ke2017lightgbm}
\Name{Ke G., Meng Q., Finley T., Wang T., Chen W., Ma W., Ye Q. \and Liu T.-Y.}
  \REVIEW{Advances in neural information processing systems}{30}{2017}{}.

\bibitem{friedman01}
\Name{Friedman J.~H.} \REVIEW{The Annals of Statistics}{29}{2001}{1189}.

\bibitem{kotsiantis13}
\Name{Kotsiantis S.~B.} \REVIEW{Artificial Intelligence Review}{39}{2013}{261}.

\bibitem{berthier07}
\Name{Berthier L. \and Jack R.~L.} \REVIEW{Physical Review E}{76}{2007}{41509}.

\bibitem{widmercooper07}
\Name{Widmer-Cooper A. \and Harrowell P.} \REVIEW{The Journal of Chemical
  Physics}{126}{2007}{154503}.

\bibitem{PhysRevE.96.062608}
\Name{Liluashvili A., \'Onody J. \and Voigtmann T.} \REVIEW{Phys. Rev.
  E}{96}{2017}{062608}.

\bibitem{RevModPhys.88.045006}
\Name{Bechinger C., Di~Leonardo R., L\"owen H., Reichhardt C., Volpe G. \and
  Volpe G.} \REVIEW{Rev. Mod. Phys.}{88}{2016}{045006}.

\bibitem{Richard2020}
\Name{Richard D., Ozawa M., Patinet S., Stanifer E., Shang B., Ridout S.~A., Xu
  B., Zhang G., Morse P.~K., Barrat J.-L., Berthier L., Falk M.~L., Guan P.,
  Liu A.~J., Martens K., Sastry S., Vandembroucq D., Lerner E. \and Manning
  M.~L.} \REVIEW{Phys. Rev. Materials}{4}{2020}{113609}.

\bibitem{ridout22}
\Name{Ridout S.~A., Rocks J.~W. \and Liu A.~J.} \REVIEW{Proceedings of the
  National Academy of Sciences}{119}{2022}{e2119006119}.

\bibitem{lundberg2017unified}
\Name{Lundberg S.~M. \and Lee S.-I.} \Book{A unified approach to interpreting
  model predictions} in proc. of \Book{Proceedings of the 31st international
  conference on neural information processing systems} 2017 pp. 4768--4777.

\end{thebibliography}

\end{document}